\documentclass[11pt,journal,onecolumn]{IEEEtran}

\usepackage{amsmath,amssymb}
\usepackage{booktabs}
\usepackage{graphicx,color,epsfig}

\title{Towards Enabling Broadband for a Billon Plus Population with TV White
Spaces}

\author{Animesh Kumar, Abhay Karandikar, Gaurang Naik, Meghna Khaturia, Shubham
Saha, Mahak Arora, and Jaspreet Singh\\
Department of Electrical Engineering\\
Indian Institute of Technology Bombay\\
Mumbai 400076 India.\\
Contact Email: animesh@ee.iitb.ac.in}

\begin{document}

\maketitle

\begin{abstract}

One of the major impediments to providing broadband connectivity in
semi-urban and rural India is the lack of robust and affordable
backhaul.  Fiber connectivity in terms of backhaul that is being
planned (or provided) by the Government of India would reach only till
rural offices (named Gram Panchayat) in the Indian rural areas.  In
this exposition, we articulate how TV white space can address the challenge
in providing broadband connectivity to a billion plus population
within India. The villages can form local Wi-Fi clusters.  The problem
of connecting the Wi-Fi clusters to the optical fiber points can be
addressed using a TV white space based backhaul (middle-mile) network. 

The amount of TV white space present in India is very large when
compared with the developed world. Therefore, we discuss a backhaul
architecture for rural India, which utilizes TV white spaces.  We also
showcase results from our TV white space testbed, which support the
effectiveness of backhaul by using TV white spaces. Our testbed
provides a broadband access network to rural population in thirteen
villages.The testbed is deployed over an area of $25$km$^2$, and
extends seamless broadband connectivity from optical fiber locations
or Internet gateways to remote (difficult to connect) rural regions.
We also discuss standards and TV white space regulations, which are
pertinent to the backhaul architecture mentioned above.

\end{abstract}

\begin{IEEEkeywords}

Wireless networks, wireless mesh networks, access networks, TV white spaces.

\end{IEEEkeywords}

\IEEEpeerreviewmaketitle

\section{Introduction}
\label{sec:intro}

In the past decades, India has witnessed ever-increasing wireless
telecom connectivity. There are over 940 million wireless telecom
subscribers with a tele-density of over 77, and India is the second
largest telecom market in the world.  These numbers along with the
advent of 3G and 4G systems would hint that wireless broadband access
in India has been solved; however, the reality is far from it! The
number of broadband subscribers is around 86 million, and the rural
tele-density is 46 (compared against the urban tele-density of 148).
However, it can be stated emphatically that the rural or affordable
wireless broadband access is an unsolved problem in India.

Rural India has purchasing capacity since it contributes $50$\% to the
GDP of India, though it has only $1.5$\% registered broadband
connections. It appears that rural broadband area is a largely
untapped market with great potential. However, there are significant
challenges in providing broadband access in the rural areas,
including: (i) small average revenue per user as a fraction of total
revenue; (ii) high capital and operation expenditure (including
license fees); (iii) affordable backhaul which is exacerbated due to a
very large population, (iv) energy cost which is worsened by lack of
reliable power supply; and, (v) geographic accessibility issues such
as right of way problems.  To \textit{alleviate} the lack of broadband
in rural areas, Government of India has been working with the
initiative BharatNet (formerly National Optical Fiber Plan or NOFN).
Within BharatNet, which is being implemented in two phases, point of
presence (PoP)  with optical connectivity at all village offices named
\textit{Gram Panchayat} will be provided. It must be noted that
mobility, at the moment, is not a major driver for broadband; instead,
\textit{primary (fixed) broadband service is the biggest requirement
in rural India}.

According to the National Telecom Policy, definition of broadband is a
$2$Mbps connection~\cite{WPCN2011}. The current broadband
subscriber-base is around $15.35$million in India. The targets of the
Government of India are high and mighty: by $2020$, the Government of
India plans to have a broadband subscriber-base of $600$million.  When
coupled with $2$Mbps definition of broadband, a subscriber base of
$250$million in $2017$, with $250$GB/month at $2$Mbps will generate
$100$Exabytes of data per month in India alone. This is $8$x larger
than the expected global mobile traffic by $2017$!  For example,
consider the city of Mumbai, the population density is $21000$/km$^2$.
Approximately $34$\% is wetland or forest. As a result, in some areas
the population density is as high as $100000$/km$^2$! With an average
household size of $4$, and if $100$\% homes have broadband, the amount
of data generated will be $50$Gbps/km$^2$. Assuming $4$ cells with
radius $< 500$m, about $12$Gbps per cell capacity will be required.
The current wireless technologies including Long Term Evolution
(LTE)/LTE-A will be unable to address this. One of the solutions,
therefore, would be to deploy small cells such as dense Wi-Fi
hotspots. However, due to non-availability of ubiquitous fiber,
backhauling of small cells is a challenge. We propose that TV UHF band
radios can be used to backhaul these dense Wi-Fi cells. 

On the rural front, there are $250,000$ village offices named Gram
Panchayat in India. The total number of villages is $638,619$, so that
each Gram Panchayat serves about $2.56$ villages on an average. Each
village has around $4$ hamlets at the periphery on an average.  As
mentioned earlier, BharatNet is an ambitious plan to provide PoP at
Gram Panchayat (offices) with optical fiber backhaul.  Since the
villages can be at a maximum distance of a few kilometers from the
Gram Panchayat, BharatNet will allay but not solve the problem of
rural broadband in India.  In each hamlet or village, a wireless
cluster can be formed (for example, by using a Wi-Fi access point),
but backhaul of the data from access points remains a challenge.
\textit{We envisage that TV white spaces (in the UHF band) can be
utilized to backhaul data from village Wi-Fi clusters to the PoP
provided by BharatNet.}

This vision raises the following important questions---Can TV UHF
band/TV white spaces be used to solve the above-mentioned backhaul
problem? How has the TV UHF band/TV white spaces utilized in the rest
of the world? How much TV white space is available in India and how
does it compare with other countries?  What network topologies in the
TV UHF band can be exploited to solve the backhaul problem? What
results are obtained from \textit{actual experimental testbed} while
performing backhaul in the TV UHF band over sparsely populated rural
areas? These questions will be subsequently answered in this
article.\footnote{In a mobility driven setup, TV white space testbed has
been used in Microsoft Corporation, Redmond to backhaul data from a
Wi-Fi cluster in a moving shuttle~\cite{chandraMBMNWA2011}.}

\section{TV white space overview}

With rising demand for bandwidth by various applications, researchers
around the world have measured the occupancy of spectrum in different
countries.  The observations suggest that except for the spectrum
allocated to services like cellular technologies, and the industrial,
scientific and medical (ISM) bands, most of the allocated spectrum is
heavily underutilized
(c.f.~\cite{islamKOQLWLTCTTS2008,lopezUCE2009,chiangRSA2007}).  The
overall usage of the analyzed spectrum ranges from 4.54\% in Singapore
to 22.57\% in Barcelona,
Spain~\cite{islamKOQLWLTCTTS2008,lopezUCE2009}.  Licensed but
unutilized TV band spectrum is called as TV white space in the
literature~\cite{FCC08260S2008,OfcomD2009}.  These white spaces in the
TV UHF band ($470$-$590$MHz) have been of particular interest owing to
the superior propagation characteristics (from a received signal
strength standpoint)~\cite{parsonsT2000}.  Its status in the world is
reviewed next.

\subsection{TV white space in various countries}
\label{sec:tvws_review}

The amount of available TV white space varies with location and time.
The available TV white space depends on regulations such as the
protection margin given to the primary user, height above average
terrain, transmission power of secondary users, and separation of
unlicensed user from the licensed ones.  Since the actual availability
of TV white spaces varies both with location and time, operators of
secondary services are interested in the amount of available white
space. TV white space estimation has been done in countries like the
United States (US), the United Kingdom (UK), Europe, Japan, and
India~\cite{harrisonMSH2010,nekoveeQ2009,vandebeekRAMT2012,shimomuraOSA2014,naikSKKQ2014}.
For instance, in Japan, out of 40 channels, on an average, 16.67
channels (41.67\%) are available in 84.3\% of the
areas~\cite{shimomuraOSA2014}.  The available TV white space by area
in Germany, UK, Switzerland, Denmark on an average ranges between 48\%
to 63 out of the 40 TV channel bands~\cite{vandebeekRAMT2012}. It must
be noted that in these TV white space studies, the IMT-A
($698$-$806$MHz) band is also included.

FCC regulations in the US and Ofcom regulations in the UK have allowed
for secondary operations in the TV white
spaces~\cite{FCC08260S2008,OfcomD2009}.  For example, FCC regulations
declare a band as unutilized if no licensed-user (primary) signal is
detected above a threshold of $-114$dBm~\cite{FCC08260S2008}.  Under
this provision, a secondary user can use the unutilized TV spectrum
provided it does not cause harmful interference to the TV band users
and it relinquishes the spectrum when a primary user starts operation.

\subsection{Standards and technologies to address TV white spaces}

IEEE 802.11af has been designed by extending IEEE 802.11ac to support
$8$MHz channels and the use of a TV white space database to inform and
control the use of spectrum by the devices~\cite{ieee80211af2013}.
IEEE 802.11af uses carrier sense multiple access 
at both the base station and the clients to share the spectrum. The
spectral efficiency of 802.11af in a single antenna configuration
varies from $0.3$bits/s/Hz to $4.5$bits/s/Hz, resulting in a maximum
throughput of $35.6$Mbps over an $8$MHz channel. IEEE 802.11-based
technologies including 802.11af and its variants, operate at a power
level of $100$mW-$1$W, and have a range of $1$km in a last mile
setting; and, it has been shown to work up to $15$km with $10$dBi
sectoral antennas in a middle-mile setting.

IEEE 802.22 is another IEEE standard designed for enabling broadband
wireless access in TV white spaces~\cite{ieee802220211}.  IEEE 802.22
uses Orthogonal Frequency Division Multiple Access as the
medium access (MAC) layer and uses centralized scheduling of the MAC
resources by the base station. The spectral efficiency of IEEE 802.22
in a single antenna configuration varies from $0.6$ bits/s/Hz to $3.1$
bits/s/Hz, with a maximum throughput of about $19$Mbps in a single
channel.

IEEE 802.15.4M is geared towards low-rate wireless personal area
networks, with applications that include machine to machine networks.
A task group IEEE 802.15.4m is addressing which technologies should be
enabled in the TV white spaces.  IEEE 802.19 is a standard for
enabling co-existence between different technologies, with specific
focus on TV white spaces~\cite{ieee8021912014}. IEEE 802.19 defines an
architecture and protocols for enabling co-existence between different
secondary networks.  Finally, 1900.7 has been established for advanced
spectrum management and next generation radios.

In the future, we expect more technologies to be designed for
operation over TV white spaces including LTE~\cite{3gpp} and IEEE
802.11ah.  We note here that LTE technology can be adopted for TV
white space by introducing design changes to the RF transmitter (for
example, by tuning the RF section to operate in TV UHF band).

\subsection{Geolocation database and white space device access rules}

To ensure coexistence of the TV broadcasters with the secondary devices,
geolocation databases have been mandated by the regulators FCC and
Ofcom~\cite{FCC08260S2008,OfcomD2009}. All devices should have a
location accuracy of $\pm 50$m and must query a certified TV white space
database to obtain an allowable channel with associated transmit
power. The list of available (unutilized by primary) channel, the
channel access schedule for $48$hours, and the transmit power that is
allowed is provided by the TV white space database.

\subsection{TV white space availability in India}

\begin{figure}[!htb]
\centering
\includegraphics[scale=0.13]{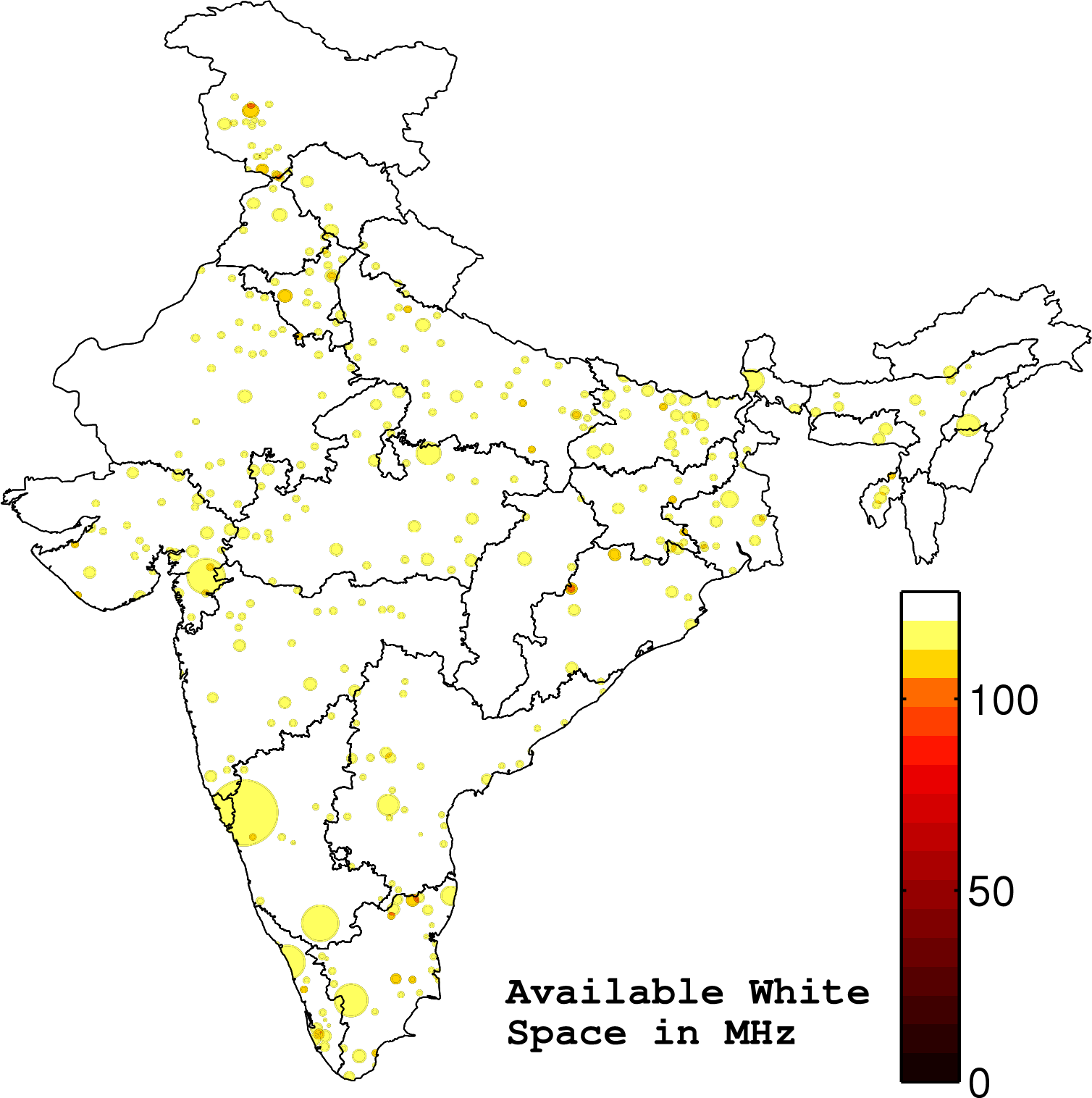}
\caption{\label{fig:fcc2} TV white space available in India is
illustrated. Observe that in almost all of the places, $100$MHz of
spectrum is nearly unutilized!}
\end{figure}

In India, the sole terrestrial TV service provider is Doordarshan
which currently transmits in two channels in most parts across India.
Currently Doordarshan has $373$ TV transmitters operating in the TV
UHF band (more precisely, TV UHF band-IV in $470$-$590$MHz) in India.
The TV UHF band consists of fifteen channels of $8$MHz each.  In
India, a small number of transmitters operate in this TV UHF bands; as
a result, apart from $8$-$16$MHz band depending on the location,
\textit{the UHF band is not utilized in India}!  Comprehensive
quantitative assessment and estimates for the TV white space in the
$470$-$590$MHz band for four zones of India have been presented by us
in the literature~\cite{naikSKKQ2014}. It has been shown that in
almost all cases at least 12 out of the 15 channels (80\%) are
available as TV white space in 100\% of the areas in
India~\cite{naikSKKQ2014} (see Fig.~\ref{fig:fcc2}).

\subsection{TV UHF band utilization in India and its policy aspects}

TV white space in India is significantly larger than other countries
reviewed above.  The common approach for TV white space utilization is
through the use of whitespace devices and associated country-specific
regulations~\cite{FCC08260S2008,OfcomD2009}.  Such white space devices
and regulations utilize the presence of database lookup, with transmit
power limitations on the unlicensed user.  While $470$-$590$MHz band has
been licensed to TV broadcasting, its usage for rural broadband can be
\textit{fundamentally different} in India than through TV white space
regulations. This fundamental difference is explained next.

In the National Frequency Allocation Plan (NFAP) of
2011~\cite{WPCN2011}, the spectrum in the frequency band
$470$-$890$MHz is earmarked for Fixed, Mobile and Broadcasting
Services.  India belongs to Region~$3$ of the ITU terrestrial spectrum
allocations. In the $470$-$590$MHz band, ITU permits \textit{Fixed,
Mobile, Broadcasting} services in the Region~3~\cite{WPCN2011}.  As
per India footnote~20~\cite{WPCN2011}, fixed services in the
$470$-$590$MHz band are allowed in India within the existing ITU
spectrum plan.  This is in contrast with Region~$1$ (including Europe)
where only broadcasting services are allowed in this band, and
Region~$2$ (including United States) where fixed services are allowed
only in $470$-$512$MHz. Accordingly, fixed services can be allowed in
$470$-$590$MHz in India. This difference accommodates high-power
transmissions by any fixed service (such as broadband base stations)
in the $470$-$590$MHz band in India.

We suggest a license-exempt registered-shared-access based regulatory
approach. Operators using TV white space spectrum may register with a
database and may have to share a channel or a sub-channel with other
users on the same channel in the vicinity. Further, different
registered operators must cooperate and coexist to ensure high average
spectral efficiency when compared to random access. Techniques such as
LBT and other suitable inter-operator co-operation techniques can be
combined with database assistance.

\section{A broadband access-network topology for rural India}

\begin{figure}[!htb]
\begin{center}
\scalebox{1.0}{\includegraphics{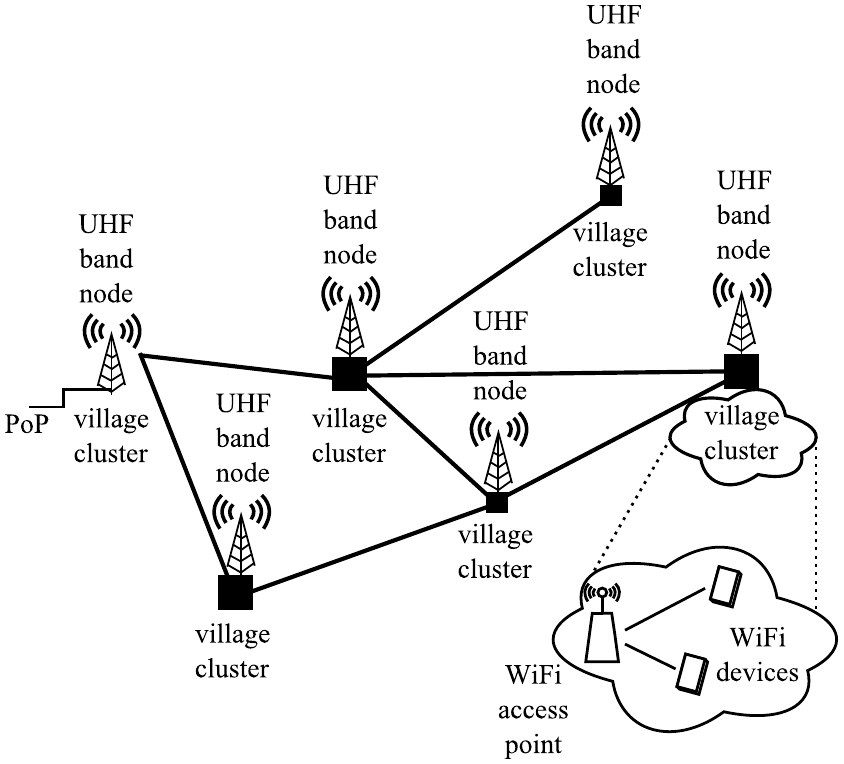}}
\end{center}
\caption{\label{fig:RuralTopology} It is assumed that there is a PoP
with a UHF band node, to provide broadband access-network in nearby
villages.  Geographically sparse and distributed villages will form
local $2.4$GHz Wi-Fi clusters with Wi-Fi access points.  Collocated
with each Wi-Fi access point, a UHF band node (a client or a base
station) will be provisioned. The data from the Wi-Fi networks will be
backhauled to the PoP by these UHF band nodes.}
\end{figure}

The $470$-$590$MHz band, \textit{henceforth the UHF band for brevity},
in India is heavily underutilized~\cite{naikSKKQ2014}, and its radio
propagation characteristics are much better than and unlicensed band
such as $2.4$GHz~\cite{parsonsT2000}. In fact, its propagation
characteristics are suitable for non-line of sight connectivity.  It
is envisaged that a broadband access network can be provided by
extending Internet coverage from a rural PoP provided by BharatNet (an
optical fiber point), by using TV white space in the UHF band.  In
such a scenario, broadband base stations operating in the UHF band
will provide \textit{backhaul} from villages to the PoP provided by
BharatNet. Each village can be served by an unlicensed-band Wi-Fi
cluster. This architecture can be used to provide affordable broadband
access-network in (rural) India, and it results in a mesh-network of
\textit{nodes} operating in the UHF band as illustrated in
Fig.~\ref{fig:RuralTopology}.
The typical distance between nodes operating in $470$-$590$MHz is
around $1$-$5$km. 

In summary, the PoP locations are provided by BharatNet. The villages
or hamlets can connect to wireless access points using $2.4$GHz Wi-Fi,
which is an affordable short-distance technology. The end-devices (for
economy of scale and for ubiquity) will connect to the UHF band mesh
network via the collocated $2.4$GHz Wi-Fi access-points.  At each
Wi-Fi access-point, a UHF band node will be provisioned. Then UHF band
nodes can be used to backhaul the data from these Wi-Fi access points
to the PoP locations.  The TV band base stations or relays can connect
in different topologies: (i) point to point; (ii) multipoint to point;
and (iii) multi-hop mesh network. Of these, we explore the most
general topology of mesh-network.

The advantages of this broadband access-network are as follows: (i)
the access technology Wi-Fi is affordable for the rural population in
India; (ii) the backhaul is cheaper due to TV UHF band propagation
characteristics (the license aspects can be handled using registered
shared access, where multiple operators can coexist through database
lookup~\cite{gurneyBEKGG2008}); and, (iii) the power consumption of
each UHF band node is low ($5$-$10$W in our testbed), so that it can
be powered through battery or solar energy. The next section
highlights the results of our testbed implemented in the multi-hop
mesh topology, in a cluster of villages near Mumbai, India.

\section{The First TV white space Test-Bed in India}

\begin{figure}[!htb]
\begin{center}
\scalebox{0.5}{\includegraphics{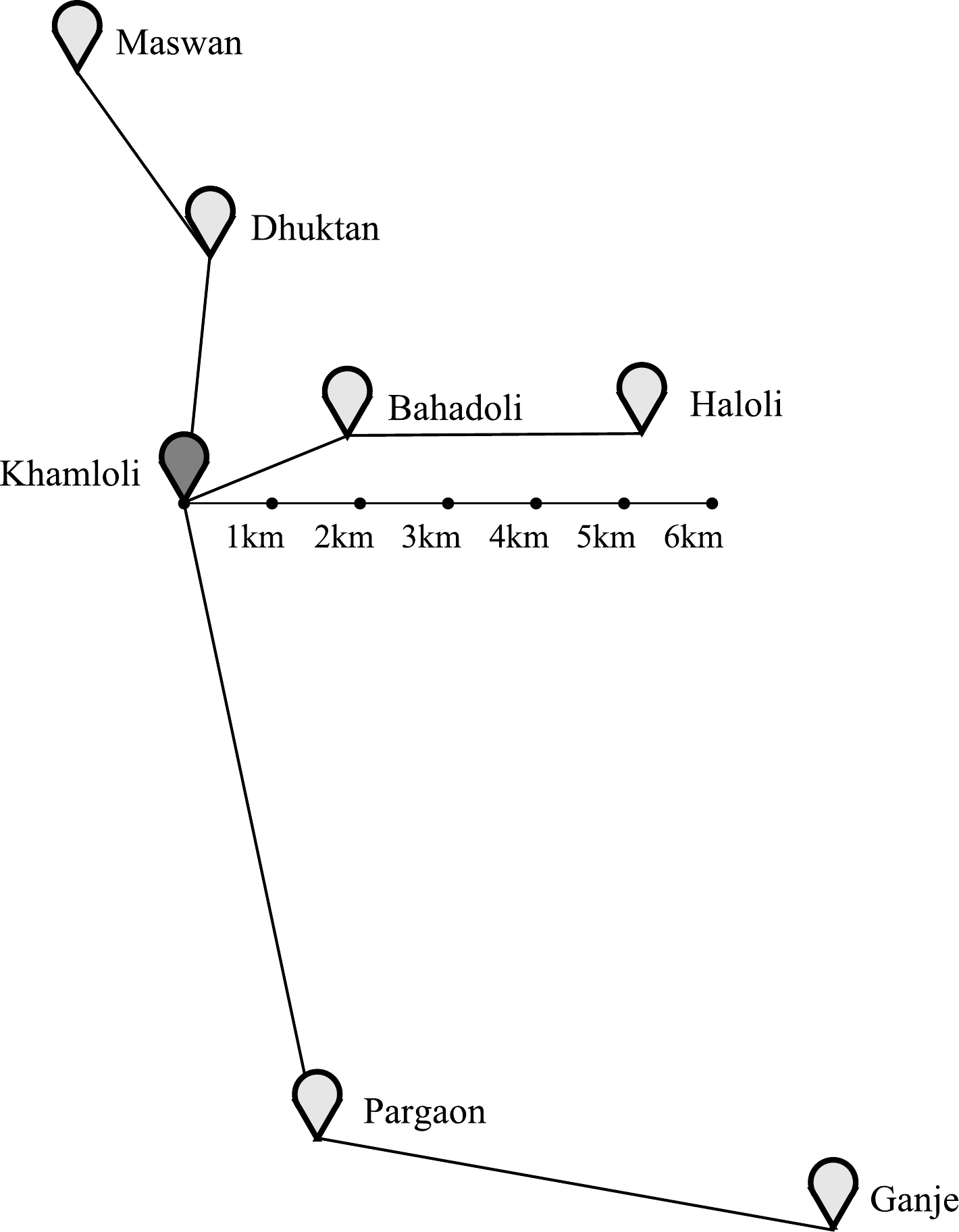}}
\end{center}
\caption{\label{fig:Palghar} An example of Indian rural topology is
illustrated.  Khamloli village has a $20$Mbps optical fiber link to
the Internet.  A wireless mesh-network can be setup with Khamloli as
the PoP to provide a broadband access-network in the nearby villages,
while leveraging on the unutilized UHF band.}
\end{figure}

An instance of the mesh-network proposed in the previous section has
been realized in a testbed at Khamloli, Maharashtra, India.  The
testbed is located in the Palghar district at about $107$km distance
from Indian Institute of Technology Bombay.  It is \textit{the first
TV white space testbed in India}.  The layout of villages surrounding
Khamloli is illustrated in Fig.~\ref{fig:Palghar}. The typical
distance between two villages is in the range of $5$km as mentioned,
and the testbed is deployed over an area of $25$km$^2$. 
%
Khamloli village node forms the PoP in this layout with a $20$Mbps
optical fiber link. With Khamloli village as PoP, a mesh-network is
setup to extend Internet from Khamloli (wirelessly in the UHF band) to
the surrounding villages.  The testbed provides a broadband access-network
to the rural population in thirteen villages or hamlets.  Our testbed
consists of $10$ UHF band nodes functioning as client and $1$ UHF
band node as base-station. The clients in UHF band are connected to
Wi-Fi hotspots to provide Internet access.

\subsection{Topographic details of the testbed}

\begin{figure*}[!htb]
\begin{center}
\scalebox{1.0}{\includegraphics{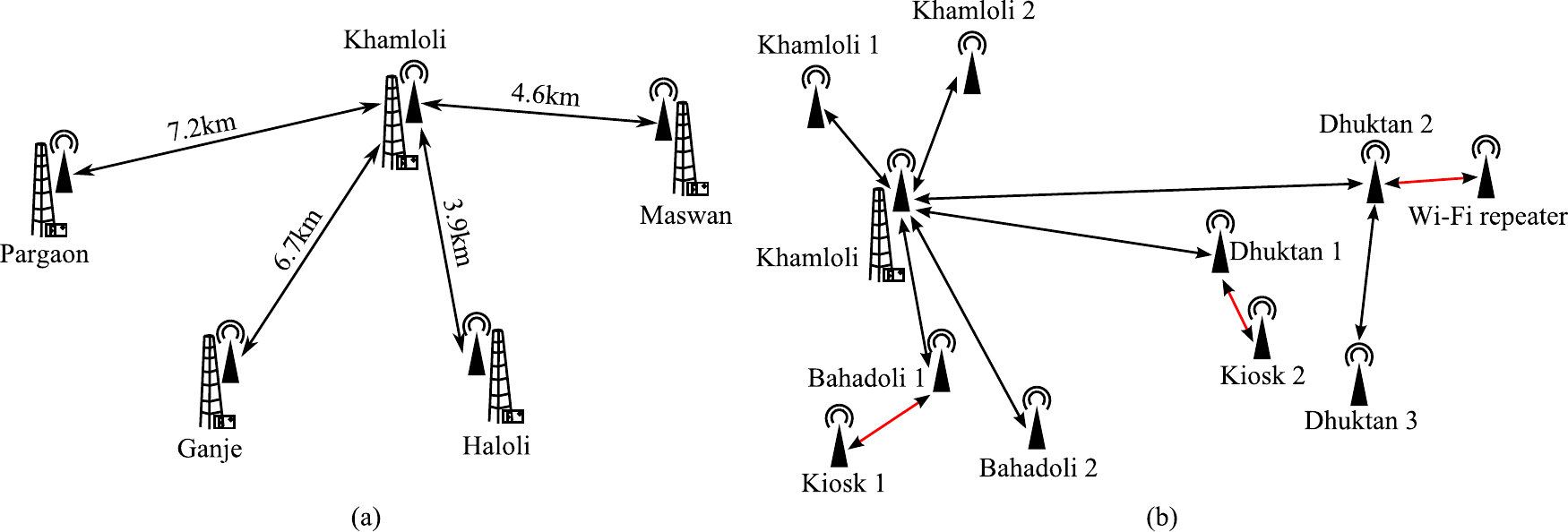}}
\end{center}
\caption{\label{fig:TestbedLayoutSDLD} Khamloli is a PoP in our
testbed. (a) The long-distance links of the testbed are illustrated.
Each tower is equipped with a UHF band base-station. (b) The
short-distance links of the testbed are illustrated. Khamloli tower
has a UHF band base-station, while other nodes have UHF band clients.}
\end{figure*}

To give an idea about the propagation characteristics of RF signal,
the topographical settings of various tower locations and UHF band
nodes are described.  It must be noted that there are significant
differences in the topographic settings between Khamloli (PoP) and the
other towers. There are four UHF band nodes setup as base-station
nodes. These result in Khamloli-Maswan, Khamloli-Haloli,
Khamloli-Ganje, and Khamloli-Pargaon (see
Fig.~\ref{fig:TestbedLayoutSDLD}(a)) links. There is no line of sight in
Khamloli-Maswan and Khamloli-Ganje links.  There is heavy vegetation
in Khamloli-Ganje link. All other links have moderate vegetation. 

There are seven UHF band nodes setup as client named Khamloli~1,
Khamloli~2, Bahadoli~1, Bahadoli~2, Dhuktan~1, Dhuktan~2, and
Dhuktan~3. These UHF band nodes are setup at a height ranging in
$4$-$6$meters, with small houses in between. Khamloli~1, Khamloli~2,
Bahadoli~1, and Bahadoli~2 have line of sight with Khamloli PoP. But,
Dhuktan~1 and Dhuktan~3 do not have line of sight while Dhuktan~2 has
partial line of sight with Khamloli PoP. Vegetation is large only in
Dhuktan~3 link (see Fig.~\ref{fig:TestbedLayoutSDLD}(b)). 

The UHF band node at Khamloli PoP is a base-station, has an
omnidirectional antenna, and is mounted at a height of $30$~meters.
At each location, a suitable site was identified for setting up the
UHF band equipment and associated accessories including power supply.
At every location marked in the testbed layout (see
Fig.~\ref{fig:TestbedLayoutSDLD}(b)), Wi-Fi routers have been
installed to test the Internet connectivity. The customer premises
equipments are Wi-Fi enabled devices such as tablets or smartphones.

\subsection{Our developed economical UHF band node prototype}

\begin{figure*}[!htb]
\begin{center}
\scalebox{1.0}{\includegraphics{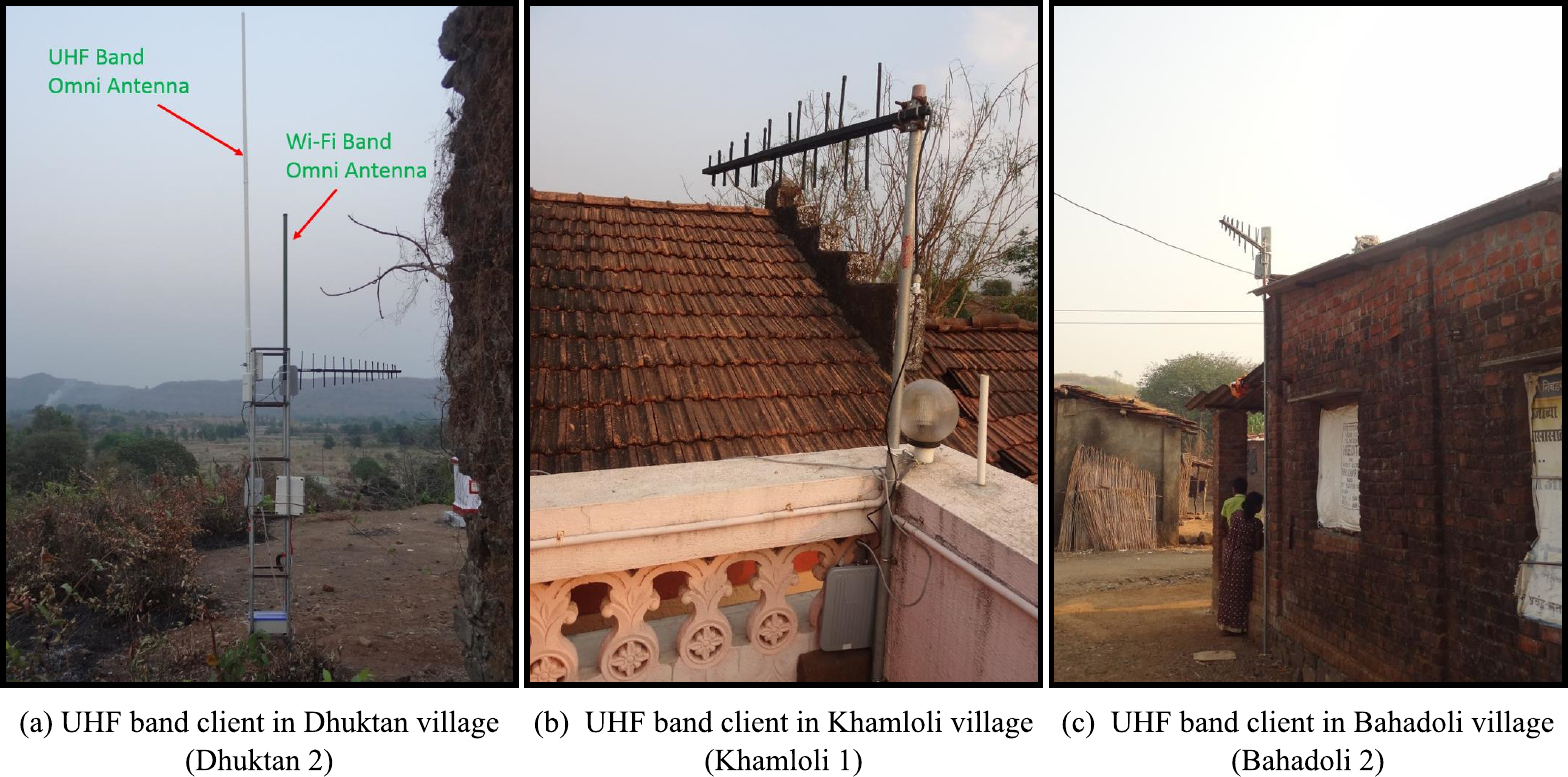}}
\end{center}
\caption{\label{fig:TwoImages} Photographs of some UHF band nodes
(clients) from Fig.~\ref{fig:TestbedLayoutSDLD} are shown. (a)
Dhuktan~2 client is mounted at the top of a hill with no houses
around. (b) and (c) UHF band clients are typically mounted at a height
of $4$-$6$meters.}
\end{figure*}

Products based on IEEE 802.22  and IEEE 802.11af standards are
available off-the shelf, but are expensive (USD$4000$-$5000$ for
base-station and USD$1000$-$2000$ for client).  From an affordable
broadband service point of view, we decided to develop a low-cost
prototype of UHF band node; and, our developed prototype costs
USD$650$ per UHF band node.  We adopted the approach of using standard
IEEE 802.11g Wi-Fi with radio-frequency of $500$MHz.  Our device has
OpenWrt operating system ported on it. This enabled us to
implement the Protocol to Access White Space (PAWS), an Internet
Engineering Task Force (IETF) standardized protocol.  \footnote{We
also experimented with IEEE 802.22 standard devices (such as those
available from Carlson Wireless), however, large scale deployment of
testbed did not use them.} Our prototype comprises of a commercially
available off-the-shelf Wi-Fi routerboard that can connect with an RF
card having a mini PCI interface.  The RF card converts the baseband
signal to $2.4$GHz, which is translated to $500$MHz (UHF band) by a
downconverter.  Few pictures from our testbed illustrating the box of
prototype are shown in Fig.~\ref{fig:TwoImages}.

\subsection{The Coexistence Handler--a UHF Band Database for India}

The testbed's UHF band devices query an OpenPAWS database setup by our
research group to select the frequency of operation, to avoid
interference to any terrestrial TV services in other geographical
regions of India~\cite{ghoshNKKO2015}.  OpenPAWS client was
implemented in OpenWrt (ported on TV UHF band node), while the
OpenPAWS server was implemented in Linux. The OpenPAWS based database
server implementation was tested to ensure its functioning with our
testbed.  An error message is generated by the OpenPAWS server in the
following three scenarios: (i) the UHF band device is outside the
regulatory domain of the TV white space database; (ii) the UHF band
device sends an AVAIL\_SPECTRUM\_REQ before initialization and
registration; or, (iii) there are no channels available at the
location of the UHF band device. In our setup, the locations of the
UHF band devices are known since it is a fixed network. This rules out
error in (i). The error in (ii) can be controlled by proper
programming. In India, it has been shown that there is a channel
available at any point~\cite{naikSKKQ2014}. Therefore, OpenPAWS
server's response will be a list of available channels with transmit
power allocation.

For the Khamloli village (PoP) in the testbed, the OpenPAWS database
declared Channel~1 to Channel~15 as available since the Channels are
available for transmission.\footnote{Note that any Channel in
$470$-$590$MHz could be used since the entire band is free in the
vicinity of Khamloli!} The transmit power was restricted to $30$dBm. After
assigning the power and channel to various UHF band nodes, the
database is updated to reflect the same.  The database created is open
for public access~\cite{ghoshNKKO2015} and can be used to view the
list of all TV towers operating in India in the UHF band along with
all operational parameters. It also displays TV towers operating on
any particular channel along with the tower's coverage area calculated
in~\cite{naikSKKQ2014}.

\subsection{Results obtained from the testbed}

The results that were obtained with our testbed after extensive
experimentation over a few months are highlighted in this section.
These comprise of throughput analysis for various links, and the
variation and latencies observed in the throughput. An ATM machine
use-case is also explained towards the end of this section. 

\begin{figure*}[!htb]
\begin{center}
\scalebox{1.0}{\includegraphics{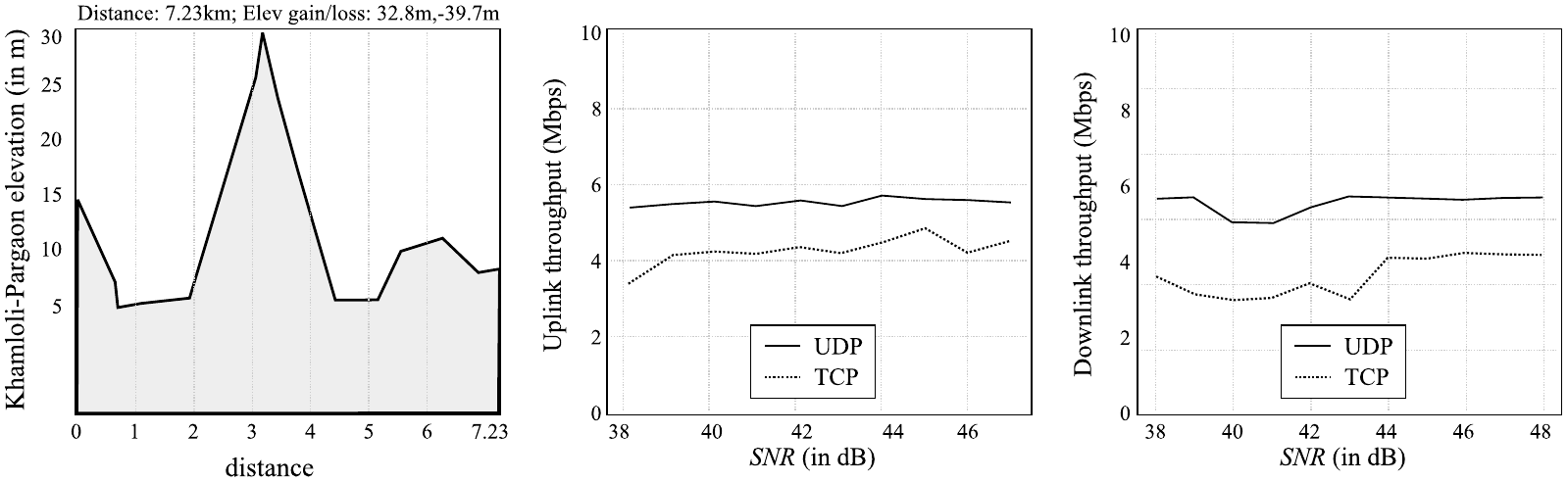}}
\end{center}
\caption{\label{fig:Plots5MHzBrief} The elevation profile of
Khamloli-Pargaon link, and the
obtained throughput for TCP/UDP protocols are shown. The channel
bandwidth was $5$MHz, and the transmit power and antenna gain were
$27$dBm and $8$dBi, respectively.}
\end{figure*}

There are many point to point links in our testbed, from which we
exemplify one.  This is a long-distance with line of sight
(Khamloli-Pargaon).  For a bandwidth of $5$MHz, the obtained uplink
and downlink throughputs (while using UDP as well as TCP) are
illustrated in Fig.~\ref{fig:Plots5MHzBrief}.  In the plots, observe
that the received signal to noise ratio (\textit{SNR}) varies since
the transmit power is kept constant (at $27$dBm with $8$dBi antenna
gain) and the receiver's distance and associated conditions vary.
Consistently a throughput of $4$-$5$Mbps (in TCP) and $5$-$6$Mbps (in
UDP) were obtained for the uplink as well as downlink.  The tool used
to monitor the network was \texttt{iperf}. TCP and UDP throughputs
over each link were computed using 500 different samples.

The variation of latencies was also examined by $25000$ randomly taken
measurements. Two extreme-range of wireless topographies were
considered---Khamloli-Ganje which represents a $6.7$km long-distance
link at $5$MHz bandwidth (large distance and low bandwidth), and
Khamloli-Dhuktan which represents a $2.3$km moderate-distance link at
$20$MHz bandwidth (small distance and large bandwidth). The latencies
for Khamloli-Ganje link varies in $2$-$15$ms, while its UDP throughput
varies in $5.6$-$8$Mbps. The latencies for Khamloli-Dhuktan link
varies in $2$-$11$ms, while its UDP throughput varies in $11$-$17$Mbps.
These results are very exciting and promising!

High speed Internet access has been provided to the villagers in
Khamloli, Bahadoli and Dhuktan using Wi-Fi hotspots deployed at 10
locations across 3 villages and 3 kiosks – one in each village.  About
60 villagers from three villages and surrounding hamlets (pada) have
been trained as `e-Sevaks' (electronic serviceman). An e-Sevak assists
other villagers in using Internet services for simple tasks like
filling college forms, paying electricity bills, and booking of train
tickets.  Kiosks set up in the three villages are run by these
e-Sevaks on a daily basis for three hours for this purpose. E-Sevaks
have been given tablets for getting familiar with these Internet
services.

An automated Teller Machine (ATM) provided by TATA Indicash is setup
at Dhuktan Gram Panchayat (village office) in order to demonstrate the
use of e-Finance capabilities of TV white space for the rural areas.
The ATM machine is on a different private LAN which is tunneled
through a Border Gateway Protocol (BGP) route from Khamloli PoP to the
ATM network. Tata Teleservices Limited and Tata Communications
provisioned the Multi-Protocol Label Switch (MPLS) leg of $64$kbps at
Khamloli, which was connected to security gateway application from
Pfsense (\texttt{http://www.pfsense.com}) by configuring one of the
optional ports as a wireless area network (WAN). The ATM machine
connects to a Wi-Fi point to point link by Ethernet; this point to
point link is connected to a Wi-Fi access point, which connects with
the PoP at Khamloli via TV white space network.

\section{Conclusions and Future Work}

In this paper, we have articulated how UHF band can address the
challenge in providing broadband connectivity to a \textit{billion
plus population} of India. As outlined in the paper, one of the major
impediments to providing broadband connectivity in semi-urban and
rural India is the lack of robust and affordable backhaul.  Even in
urban areas, one of the major impediments for widespread deployment of
Wi-Fi Hotspots is the lack of connectivity from Wi-Fi access points to
optical fiber gateways. Fiber connectivity in terms of backhaul that
is being planned (or provided) by the Government of India would reach
only till the Gram Panchayat in the rural areas.  In such a scenario,
the problem of connecting the Wi-Fi clusters to the optical fiber PoP
can be addressed using a TV white space based backhaul (middle-mile)
network.

We believe that a cost effective solution for backhaul would require a
license exempt database assisted approach for TV white space spectrum
management.  Since UHF band is sparsely utilized in India by the
broadcaster, the challenge is not primary-secondary co-existence (as
in many countries) but secondary-secondary co-existence.  Multiple
operators should be able to share the TV spectrum and co-exist. While
listen before talk (LBT) approach as in IEEE 802.11af standard is one
of the options for co-existence, it will pose challenges in rural
regions with large cell radius, due to superior propagation
characteristics of sub-GHz spectrum.  A combination of database
assisted and LBT is a topic for future investigation for providing
``primary'' broadband connectivity by possibly many local operators.

Dynamic resource allocation algorithm with fair 
sharing of resources between multiple operators (co-primary) is another 
area that requires more attention.Since TV band network is being proposed 
as middle mile network for backhauling Wi-Fi clusters,both Wi-Fi access 
points and TV band radios can be controlled through a Software Defined 
Networking (SDN) controller. This SDN controller can also be integrated 
with Database controller. Finally SDN controller can also be employed to 
dynamically configure flow routing in a more complex topology of 
multi-hop mesh based middle mile. A SDN enabled Policy based Radios 
deployed for middle mile fixed services can set the vision for 5G 
for India. Recently, the need for such vision for 5G has been also
articulated by Eriksson and van de Beek~\cite{erikssonVI2015}.

We are currently investigating these
topics as enhancements to existing TV white space standards. These
topics are also under discussion in Telecom Standards Development
Society of India (\texttt{http://www.tsdsi.org}).

\bibliographystyle{IEEEtran} 
\bibliography{../../../../references}

\end{document}